\documentclass[12pt]{iopart}
\usepackage{iopams}
\usepackage{epsf}
\usepackage{latexsym}
\usepackage{epsfig}
\usepackage{amssymb}
\usepackage{amsfonts}
\usepackage{graphicx}

\begin{document}

\title{Entropy landscape of solutions in the binary perceptron problem}

\author{Haiping Huang$^{1,2}$, K. Y. Michael Wong$^{2}$ and Yoshiyuki Kabashima$^{1}$}
\address{$^{1}$ Department of Computational Intelligence and Systems Science,
Tokyo Institute of Technology, Yokohama 226-8502, Japan}
\address{$^{2}$ Department of Physics, The Hong Kong University of Science and
Technology, Clear Water Bay, Hong Kong, China}
\date{\today}

\begin{abstract}
 The statistical picture of the solution space for a binary
 perceptron is studied. The binary perceptron learns a random
 classification of input random patterns by a set of binary synaptic weights. The learning of this
 network is difficult especially when the pattern (constraint) density is close to the capacity, which is supposed to be intimately related to the structure
 of the solution space. The geometrical organization is elucidated
 by the entropy landscape from a reference configuration and of
 solution-pairs separated by a given Hamming distance in the solution space. We
 evaluate the entropy at the annealed level as well as replica symmetric level
 and the mean field result is confirmed by the numerical simulations on single instances using the proposed message passing algorithms.
 From the first landscape (a random configuration as a reference),
 we see clearly how the solution space shrinks as more constraints
 are added. From the second landscape of solution-pairs, we deduce
 the coexistence of clustering and freezing in the solution space.

\end{abstract}

\pacs{89.75.Fb, 87.19.L-, 75.10.Nr}
 \maketitle

\section{Introduction}
Learning in a single layer of feed forward neural network with
binary synapses has been studied either based on statistical
mechanics
analysis~\cite{Krauth-1989,Gutf-1990,Barkai-1992pra,Kaba-09} or in
algorithmic
aspects~\cite{Kohler-1990,Patel-1993,Horner-1992a,Bouten-1998,Zecchina-2006,Baldassi-2007,Huang-2010jstat,Huang-2011epl}.
This network can learn an extensive number $P=\alpha N$ of random
patterns, where $N$ is the number of synapses and $\alpha$ denotes
the constraint density. The critical $\alpha$ (also called the
capacity) separating the learnable phase from unlearnable phase is
predicted to be $\alpha_{s}\simeq0.833$ where the entropy
vanishes~\cite{Krauth-1989}. A solution is defined as a
configuration of synaptic weights to implement the correct
classification of $P$ random input patterns. Above $\alpha_{s}$, no
solutions can be found with high probability (converging to $1$ in
the thermodynamic limit). The replica symmetric solution presented
in Ref.~\cite{Krauth-1989} has been shown to be stable up to the
capacity, which is in accordance with the convexity of the solution
space~\cite{Barkai-1992pra}. Note that the solutions disappear at
the threshold $\alpha_{s}$ still maintaining a typical finite value
of Hamming distance between them, which is quite distinct from the
case in the continuous perceptron with real-valued synapses. In the
continuous perceptron, this distance tends to zero when the
solutions disappear at the corresponding
threshold~\cite{Gardner-1988a}. On the other hand, many local search
algorithms~\cite{Patel-1993,Horner-1992a,Huang-2010jstat,Huang-2011epl}
were proposed to find solutions of the perceptron learning problem,
however, the search process slows down with increasing $\alpha$, and
the critical $\alpha$ for the local search
algorithm~\cite{Huang-2011epl} decreases when the number of synapses
increases. This typical behavior of the stochastic local search
algorithm is conjectured to be related to the geometrical
organization of the solution space~\cite{Horner-1992a,Kaba-09}. In
order to acquire a better understanding for the failure of the local
search strategy, we compute the entropy landscape both from a
reference configuration and for solution-pairs with a given distance
in the solution space. Both distance landscapes contain rich
information about the detailed structure of the solution space and
then can help us understand the observed glassy behavior of the
local search algorithms. Throughout the paper, the term distance
refers to the Hamming distance. The distance landscape has been well
studied in random constraint satisfaction problems defined on
diluted or sparse random
graphs~\cite{Mora-2006jstat,Daude-2008,Zecchina-2008entropy,ZH-2010jstat}.

\begin{figure}
\centering
         \includegraphics[bb=119 561 469 745,scale=0.6]{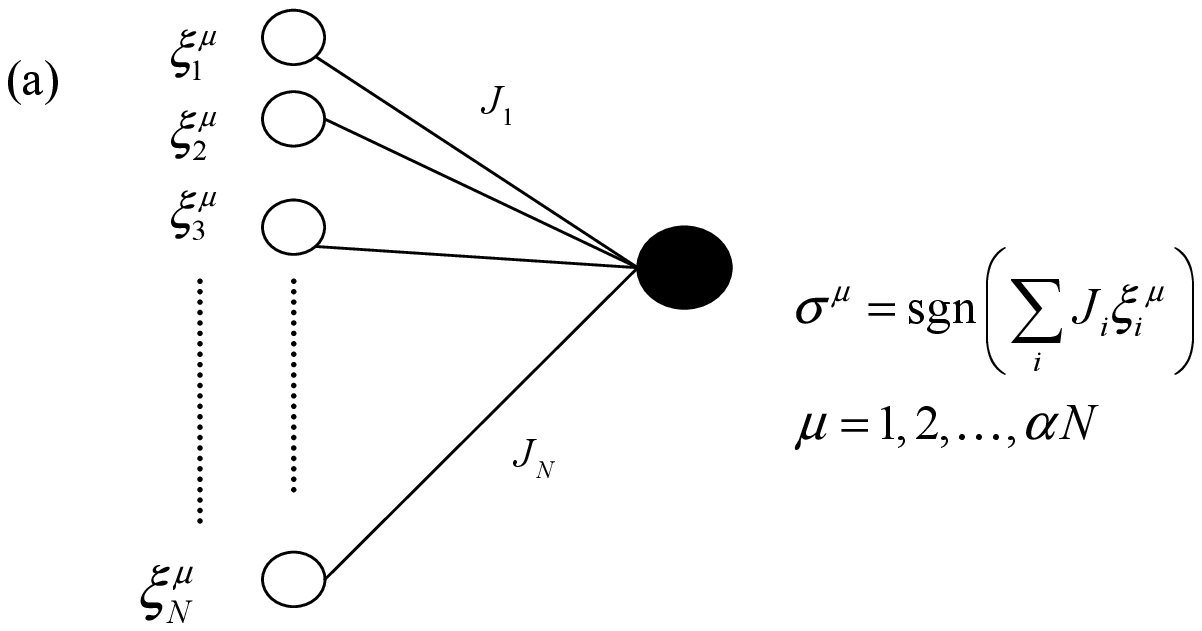}
     \hskip .5cm
     \includegraphics[bb=98 594 262 747,scale=0.8]{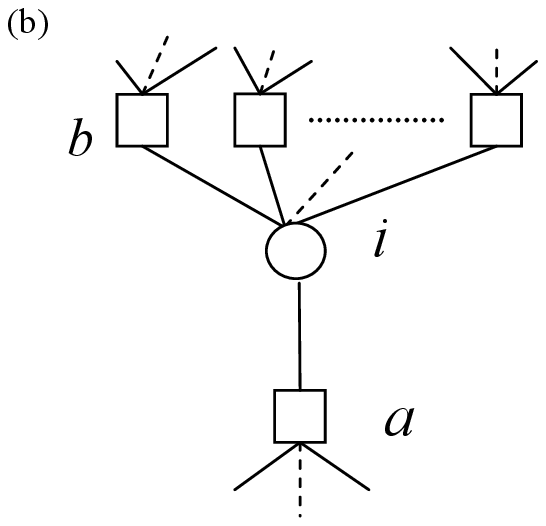}\vskip .2cm
  \caption{
     The sketch of the binary perceptron and the factor graph representation. (a) $N$ input units (open circles)
     feed directly to a single output unit (solid circle). A binary input pattern
    $(\xi_1^\mu, \xi_2^\mu, \ldots, \xi_N^\mu)$ of length $N$
    is mapped through a sign function to a binary output $\sigma^{\mu}$, i.e.,
    $\sigma^\mu = {\rm sgn}\bigl(\sum_{i=1}^{N} J_{i}\xi_{i}^{\mu}\bigr)$. The set of $N$
      binary synaptic weights $\{J_{i}\}$ is regarded as a solution of the perceptron
     problem if the output $\sigma^\mu= \sigma_0^\mu$ for each of the $P= \alpha N$ input patterns $\mu \in [1, P]$, where
     $\sigma_0^\mu$ is a preset binary value.
      (b) Each circle denotes the variable node whose value takes $J_{i}$ with $i$ its index. The square is the function node denoting a
      random binary pattern to be learned. If the pattern is learned by the synaptic vector $\boldsymbol{J}$, the value of the corresponding function node takes
      zero. The dotted line
      represents other $P-4$ function nodes while the dashed line
      for variable node $i$ means $i$ is connected to other $P-4$
      function nodes and that for function node means the function node (e.g., $b$)
      is connected to other $N-3$ variable nodes.
   }\label{perc}
 \end{figure}

Learning in the binary perceptron can be mapped onto a bipartite
graph where variable node represents synaptic weight and function
node represents the input random pattern to be learned (see
figure~\ref{perc} (b)). This graph is also called graphical model or
factor graph~\cite{Frey-2001}. The efficient message passing
learning algorithm for the binary perceptronal learning problem has
been derived using the cavity method and this factor graph
representation~\cite{Zecchina-2006}. In this paper, we focus on the
typical property of the solution space in random ensembles of the
binary perceptronal learning problem. We apply the replica trick
widely used to study disordered systems~\cite{Mezard-1987} to
compute the statistical properties in the thermodynamic limit. To
confirm the mean field result computed using the replica approach,
we derive the message passing equations in the cavity context which
can be applied on single random instances of the current problem. In
this context, we apply the decorrelation assumption as well as the
central limit theorem to derive the formula at the replica symmetric
level. This assumption arises from the weak correlation among
synaptic weights (within one pure state~\cite{Mezard-1987}) and
among input patterns~\cite{cavity-1989}. The efficiency of the
inspired message passing algorithms in loopy systems has been
observed in
Refs.~\cite{Kaba-2003a,Neirotti-2005epl,Zecchina-2006,Montanari-2011}
while the underlying mechanism still needs to be fully understood.
However, our cavity method focuses on the physical content and
yields the same result as that obtained using replica
approach~\cite{cavity-1989,Wong-1995epl,Barra-2006}.

The remainder of this paper is organized as follows. The random
classification by the binary perceptron is defined in
Sec.~\ref{sec_bperc}. In Sec.~\ref{sec_sdr}, we derive the
self-consistent equations to compute the distance landscape (entropy
landscape) from a reference configuration, i.e., to count the number
of solutions at a distance from the reference configuration. Both
the annealed and replica symmetric (RS) computations of this entropy
landscape are presented. We also derive the message passing
equations for single instances in this section using the cavity
method and factor graph representation. In Sec.~\ref{sec:sdp}, the
landscape of Hamming distances between pairs of solutions is
evaluated at both annealed approximation and RS ansatz, and the
associated message passing equations are proposed as well.
Discussion and conclusion are given in Sec.~\ref{sec_con}.

\section{Problem definition}
\label{sec_bperc}

The binary perceptron realizes a random classification of $P$ random
input patterns (see figure~\ref{perc}(a)). To be more precise, the
learning task is to find an optimal set of binary synaptic weights
(solution) $\{J_{i}\}_{i=1}^{N}$ that could map correctly each of
random input patterns $\{\xi_{i}^{\mu}\}(\mu=1,\ldots,P)$ to the
desired output $\sigma_{0}^{\mu}$ which is assigned a value $\pm1$
at random. $P$ is proportional to $N$ with the coefficient $\alpha$
defining the constraint density (each input pattern serves as a
constraint for all synaptic weights, see figure~\ref{perc} (b)). The
critical value is $\alpha_{s}\simeq0.833$ below which the solution
space is non-empty~\cite{Krauth-1989}. Given the input pattern
$\boldsymbol{\xi}^{\mu}$, the actual output $\sigma^{\mu}$ of the
perceptron is $\sigma^{\mu}={\rm
sgn}\left(\sum_{i=1}^{N}J_{i}\xi_{i}^{\mu}\right)$ where $J_{i}$
takes $\pm1$ and $\xi_{i}^{\mu}$ takes $\pm1$ with equal
probabilities. If $\sigma^{\mu}=\sigma_{0}^{\mu}$, we say that the
synaptic weight vector $\boldsymbol{J}$ has learned the $\mu$-th
pattern. Therefore we define the number of patterns mapped
incorrectly as the energy cost
\begin{equation}\label{cost}
E(\boldsymbol{J})=\sum_{\mu}\Theta\left(-\frac{\sigma_{0}^{\mu}}{\sqrt{N}}\sum_{i=1}^{N}J_{i}\xi_{i}^{\mu}\right)
\end{equation}
where $\Theta(x)$ is a step function with the convention that
$\Theta(x)=0$ if $x\leq0$ and $\Theta(x)=1$ otherwise. The prefactor
$N^{-1/2}$ is introduced to ensure that the argument of the step
function remains at the order of unity, for the sake of the
following statistical mechanical analysis in the thermodynamic
limit. In the current setting, both $\{\xi_{i}^{\mu}\}$ and the
desired output $\{\sigma_{0}^{\mu}\}$ are generated randomly
independently. Without loss of generality, we assume
$\sigma_{0}^{\mu}=+1$ for any input pattern in the remaining part of
this paper, since one can perform a gauge transformation
$\xi_{i}^{\mu}\rightarrow\xi_{i}^{\mu}\sigma_{0}^{\mu}$ to each
input pattern without affecting the result.

\section{Distance landscape from a reference configuration}
\label{sec_sdr}

In this section, we consider the entropy landscape from a reference
configuration (which is not a solution). This entropy counts the
number of solutions at a distance $Nd$ from the reference
configuration $\boldsymbol{J^{*}}$. The behavior of this entropy
landscape reflects the geometrical organization of the solution
space. Since we concentrate on the ground state ($E=0$), we take the
inverse temperature $\beta\rightarrow\infty$ and introduce a
coupling field $x$ to control the distance between solutions and the
reference configuration. The partition function for this setting is
\begin{equation}\label{PTF}
Z=\sum_{\boldsymbol{J}}\prod_{\mu}\Theta\left(\frac{1}{\sqrt{N}}\sum_{i}J_{i}\xi_{i}^{\mu}\right)\exp\left[x\sum_{i}J_{i}J_{i}^{*}\right]
\end{equation}
where the sum $\sum_{\boldsymbol{J}}$ goes over all possible
synaptic weight vectors and $\sum_{i}$ means the sum over all variable
nodes. Under the definition of the overlap
$\tilde{q}\equiv\frac{1}{N}\sum_{i}J_{i}J_{i}^{*}$, the partition function
can be written as
\begin{equation}\label{PTF02}
Z=\sum_{\tilde{q}}\exp\left[N(s(\tilde{q})+x\tilde{q})\right]
\end{equation}
where $e^{Ns(\tilde{q})}$ is the number of solutions with the overlap $\tilde{q}$.
In the thermodynamic limit $N\rightarrow\infty$, the saddle point
analysis leads to $f(x)\equiv\frac{1}{N}\log
Z=\max_{\tilde{q}}\left[s(\tilde{q})+x\tilde{q}\right]$ where $f(x)$ is defined as the free
energy density. Therefore, we can determine the entropy $s(\tilde{q})$ by a
Legendre transform~\cite{Zecchina-2008entropy,ZH-2010jstat}
\begin{eqnarray}\label{Legendre01}
s(\tilde{q})&=&\min_{x}\left[f(x)-x\tilde{q}\right],\\
\tilde{q}(x)&=&\frac{df(x)}{dx}\label{Legendre02}
\end{eqnarray}
where $\tilde{q}$ is related to $d$ through $d=\frac{1-\tilde{q}}{2}$ and then the
entropy density can be expressed as a function of the distance $d$
which can be understood as the probability that a synaptic weight
takes different values in $\boldsymbol{J}$ and $\boldsymbol{J^{*}}$.
One recovers the total number of solutions by setting $x=0$ in
Eq.~(\ref{PTF}).

\subsection{Annealed approximation for $s(d)$}
\label{subsec:ann01}

We first calculate the annealed entropy density $s_{\rm{ann}}(d)$
which serves as the upper bound (Jensen's
inequality~\cite{Cover-1991}) for the true value of the entropy
density. Actually, the free energy $\log Z$ should be averaged over
the random input patterns. However, the annealed approximation
alternatively performs the average of the partition function first
and then takes the logarithmic operation as
$f_{\rm{ann}}\equiv\frac{1}{N}\log\left<Z\right>$ where the average
is taken over the distribution of the random input patterns. This
can be computed as
\begin{eqnarray}\label{annZ01}
\left<Z\right>&=&\left<\sum_{\boldsymbol{J}}\prod_{\mu}\Theta\left(\frac{1}{\sqrt{N}}\sum_{i}J_{i}\xi_{i}^{\mu}\right)\exp\left[x\sum_{i}J_{i}J_{i}^{*}\right]\right>
\nonumber\\
&=&\int d\tilde{q}\int\frac{d\hat{q}}{2\pi
\mathrm{i}/N}\exp\left[N\left(-\hat{q}\tilde{q}+x\tilde{q}-\alpha\log2+\log(2\cosh\hat{q})\right)\right]
\end{eqnarray}
where the integral representation of $\Theta(\cdot)$ is used and the
conjugated counterpart $\hat{q}$ of the overlap $\tilde{q}$ is
introduced as a Dirac delta function
$\delta\left(\tilde{q}-\frac{1}{N}\sum_{i}J_{i}J_{i}^{*}\right)$ is
inserted~\cite{Engel-2001}. A saddle point analysis results in
\begin{equation}\label{annf01}
f_{\rm{ann}}=\max_{\tilde{q},\hat{q}}\{-\hat{q}\tilde{q}+x\tilde{q}-\alpha\log2+\log(2\cosh\hat{q})\}
\end{equation}
where the saddle point equation reads $\hat{q}=x,\tilde{q}=\tanh\hat{q}$.
Using Eq.~(\ref{Legendre01}) and the saddle point equation, we get
the annealed entropy density
\begin{equation}\label{annS01}
s_{\rm{ann}}(d)=-\alpha\log2-d\log d-(1-d)\log(1-d).
\end{equation}

\subsection{Replica symmetric computation of $s(d)$}
\label{subsec:RS01}

The free energy density $f(x)$ is a self-averaging quantity whose
value concentrates in probability around its expectation in the
thermodynamic limit~\cite{Guerra-2002}, and its average over the
random input patterns is very difficult to compute because the
logarithm appears inside the average. The replica trick bypasses
this difficulty by using the identity $\log Z=\lim_{n\rightarrow
0}\frac{Z^{n}-1}{n}$. Then the disorder averaged free energy density
can be computed by first averaging an integer power of the partition
function and then letting $n\rightarrow 0$ as
\begin{equation}\label{repf01}
f=\lim_{n\rightarrow 0,
N\rightarrow\infty}\frac{\log\left<Z^{n}\right>}{nN}.
\end{equation}
Although the replica method is not generally rigorous, the obtained
theoretical result can be checked by numerical simulations. To
compute $\left<Z^{n}\right>$, we introduce $n$ replicated synaptic
weight vectors $\boldsymbol{J}^{a}$($a=1,\ldots,n$) as follows.
\begin{eqnarray}\label{repf02}
\fl\left<Z^{n}\right>=\left<\sum_{\{\boldsymbol{J}^{a}\}}\prod_{a,\mu}\Theta\left(\frac{1}{\sqrt{N}}\sum_{i}J^{a}_{i}\xi_{i}^{\mu}\right)
\exp\left[x\sum_{i,a}J^{a}_{i}J_{i}^{*}\right]\right>\nonumber\\
\fl=\int\prod_{a<b}\frac{dq^{ab}d\hat{q}^{ab}}{2\pi
\mathrm{i}/N}\exp\left[-N\sum_{a<b}q^{ab}\hat{q}^{ab}+N\alpha\log
G_{0}(\{q^{ab}\})+NG_{1}(\{\hat{q}^{ab}\})\right],
\end{eqnarray}
where $G_{0}$ and $G_{1}$ are expressed respectively as
\begin{eqnarray}\label{G0G1}
G_{0}(\{q^{ab}\})&=&\prod_{a}\left[\int\frac{d\lambda^{a}}{2\pi}\int_{0}^{\infty}dt^{a}\right]e^{\mathrm{i}\sum_{a}\lambda^{a}t^{a}-\sum_{a<b}\lambda^{a}\lambda^{b}q^{ab}-\frac{1}{2}\sum_{a}(\lambda^{a})^{2}},\\
G_{1}(\{\hat{q}^{ab}\})&=&\log\sum_{J^{a}:a=1,\ldots,n}e^{\sum_{a<b}\hat{q}^{ab}J^{a}J^{b}+x\sum_{a}J^{a}J^{*}},
\end{eqnarray}
where we have introduced the replica overlap
$q^{ab}\equiv\frac{1}{N}\sum_{i}J_{i}^{a}J_{i}^{b}$ and its
associated conjugated counterpart $\hat{q}^{ab}$. The replica
symmetric ansatz assumes $q^{ab}=q, \hat{q}^{ab}=\hat{q}$ for $a\neq
b$. Now using the saddle point analysis, we finally arrive at the formula of the free energy density and the corresponding saddle
point equations,
\begin{eqnarray}\label{repf03}
f(x)&=&\frac{\hat{q}}{2}(q-1)+\alpha\int Dz\log
H\left(\sqrt{\frac{q}{1-q}}z\right)\nonumber\\
&+&\int
Dz\log\left[2\cosh(\sqrt{\hat{q}}z+x)\right],\\
q&=&\int Dz\tanh^{2}(\sqrt{\hat{q}}z+x),\label{repf03a}\\
\hat{q}&=&\frac{\alpha}{1-q}\int
Dz\left[G\left(\sqrt{\frac{q}{1-q}}z\right)/H\left(\sqrt{\frac{q}{1-q}}z\right)\right]^{2},\label{repf03b}
\end{eqnarray}
where $G(x)=\exp(-x^{2}/2)/\sqrt{2\pi}$ and
$H(x)\equiv\int_{x}^{\infty} Dz$ with the Gaussian measure $Dz\equiv
G(z)dz$. After the fixed point of the self-consistent
equations~(\ref{repf03a}) and~(\ref{repf03b}) is obtained, the
entropy landscape $s(d)$ is computed as
\begin{equation}\label{sdrep}
s(x)=f(x)-x\int Dz\tanh(\sqrt{\hat{q}}z+x).
\end{equation}
Note that the final expression of $s(x)$ does not depend on the
reference configuration and the integral in the second term of
Eq.~(\ref{sdrep}) is $\tilde{q}(x)$ defined in Eq.~(\ref{Legendre02}).

\subsection{Message passing equations for single instances}
\label{subsec:MPA01}

 In this section, we derive the message passing equations to compute the entropy landscape for single instances under the replica symmetric ansatz. To derive the
self-consistent equation, we apply the cavity
method~\cite{cavity-1989,Zecchina-2006} and first define two kinds
of cavity probabilities. One is the probability $p_{i\rightarrow
a}^{J_{i}}$ that variable node $i$ in figure~\ref{perc} (b) takes
value $J_{i}$ in the absence of constraint $a$. The other is
$\hat{p}_{b\rightarrow i}^{J_{i}}$ staying for the probability that
constraint $b$ is satisfied (pattern $\mu=b$ is learned) if synaptic
weight $i$ takes $J_{i}$. According to the above definitions, the
self-consistent equation for these two kinds of probabilities is
readily obtained as
\begin{eqnarray}\label{bp0}
p_{i\rightarrow a}^{J_{i}}&=&\frac{1}{Z_{i\rightarrow
a}}e^{xJ_{i}J_{i}^{*}}\prod_{b\in\partial i\backslash
a}\hat{p}_{b\rightarrow i}^{J_{i}},\\
\hat{p}_{b\rightarrow i}^{J_{i}}&=&\sum_{\{J_{j},j\in\partial
b\backslash
i\}}\Theta\left(\frac{1}{\sqrt{N}}\sum_{j}J_{j}\xi_{j}^{b}\right)\prod_{j\in\partial
b\backslash i}p_{j\rightarrow b}^{J_{j}},\label{bp1}
\end{eqnarray}
where $Z_{i\rightarrow a}$ is a normalization constant, $\partial
i\backslash a$ denotes the neighbors of node $i$ except constraint
$a$ and $\partial b\backslash i$ denotes the neighbors of constraint
$b$ except variable node $i$. Eqs.~(\ref{bp0}) and~(\ref{bp1}) are
actually the belief propagation
equations~\cite{cavity-1989,Zecchina-2006}. For the binary
perceptron, directly solving the belief propagation equations is
impossible. To reduce the computational complexity, we define
$w_{b\rightarrow i}\equiv\frac{1}{\sqrt{N}}\sum_{j\neq
i}J_{j}\xi_{j}^{b}$. Note that the sum involves $N-1$ independent
random terms, as a result, the central limit theorem implies that
$w_{b\rightarrow i}$ follows a Gaussian distribution with mean
$\left<w_{b\rightarrow i}\right>$ and variance
$\sigma_{w_{b\rightarrow i}}^{2}$ where $\left<w_{b\rightarrow
i}\right>=\frac{1}{\sqrt{N}}\sum_{j\neq i}m_{j}\xi_{j}^{b}$ and
$\sigma_{w_{b\rightarrow i}}^{2}=\frac{1}{N}\sum_{j\neq
i}(1-m_{j}^{2})$. Within the RS ansatz, the clustering property
$\left<J_{i}J_{j}\right>-\left<J_{i}\right>\left<J_{j}\right>\simeq
0$ for $i\neq j$ in the thermodynamic limit is used to get the
variance~\cite{Zecchina-2006}. $m_{j}\equiv\left<J_{j}\right>$ is
the magnetization in statistical physics language. By separating the
term $\frac{1}{\sqrt{N}}J_{i}\xi_{i}^{b}$ from the sum in the
$\Theta(\cdot)$ of Eq.~(\ref{bp1}), and approximating the sum
$\sum_{\{J_{j},j\in\partial b\backslash i\}}$ by an integral over
$w_{b\rightarrow i}$, we get finally
\begin{equation}\label{Phat}
\hat{p}_{b\rightarrow
i}^{J_{i}}=H\left(-\frac{J_{i}\xi_{i}^{b}+\hat{w}_{b\rightarrow
i}}{\sqrt{\hat{\sigma}_{b\rightarrow i}}}\right)
\end{equation}
where $\hat{w}_{b\rightarrow i}=\sum_{j\in\partial b\backslash
i}m_{j\rightarrow b}\xi_{j}^{b}$ and $\hat{\sigma}_{b\rightarrow
i}=\sum_{j\in\partial b\backslash i}(1-m_{j\rightarrow b}^{2})$ in
which the cavity magnetization $m_{j\rightarrow b}\equiv\tanh
h_{j\rightarrow b}$. Using Eqs.~(\ref{bp0}) and~(\ref{Phat}), the
cavity field $h_{j\rightarrow b}$ can be obtained in the
log-likelihood representation
\begin{eqnarray}\label{cavH01}
\fl h_{j\rightarrow b}&=&\frac{1}{2}\log\frac{p_{j\rightarrow
b}^{+1}}{p_{j\rightarrow b}^{-1}}=xJ_{j}^{*}+\sum_{a\in\partial j\backslash b}u_{a\rightarrow j},\\
\fl u_{a\rightarrow j}&=&\frac{1}{2}\log\frac{\hat{p}_{a\rightarrow
j}^{+1}}{\hat{p}_{a\rightarrow j}^{-1}}=\frac{1}{2}\left[\log
H\left(-\frac{\xi_{j}^{a}+\hat{w}_{a\rightarrow
j}}{\sqrt{\hat{\sigma}_{a\rightarrow j}}}\right)-\log
H\left(\frac{\xi_{j}^{a}-\hat{w}_{a\rightarrow
j}}{\sqrt{\hat{\sigma}_{a\rightarrow
j}}}\right)\right].\label{cavH02}
\end{eqnarray}
Notice that the cavity bias $u_{a\rightarrow j}$ can be approximated
by $\frac{\xi_{j}^{a}G\left(\frac{\hat{w}_{a\rightarrow
j}}{\sqrt{\hat{\sigma}_{a\rightarrow
j}}}\right)}{H\left(-\frac{\hat{w}_{a\rightarrow
j}}{\sqrt{\hat{\sigma}_{a\rightarrow
j}}}\right)\sqrt{\hat{\sigma}_{a\rightarrow j}}}$ in the large $N$
limit. Eqs.~(\ref{cavH01}) and~(\ref{cavH02}) constitute the
recursive equations to compute the free energy density in the Bethe
approximation~\cite{MM-2009}
\begin{eqnarray}\label{fRS01}
f(x)&=&\frac{1}{N}\sum_{i}\Delta f_{i}-\frac{N-1}{N}\sum_{a}\Delta
f_{a},\\
\Delta f_{i}&=&\log\left[e^{xJ_{i}^{*}}\prod_{b\in\partial
i}H\left(-\frac{\xi_{i}^{b}+\hat{w}_{b\rightarrow
i}}{\sqrt{\hat{\sigma}_{b\rightarrow
i}}}\right)+e^{-xJ_{i}^{*}}\prod_{b\in\partial
i}H\left(\frac{\xi_{i}^{b}-\hat{w}_{b\rightarrow
i}}{\sqrt{\hat{\sigma}_{b\rightarrow i}}}\right)\right],\\
\Delta f_{a}&=&\log
H\left(-\frac{\hat{w}_{a}}{\sqrt{\hat{\sigma}_{a}}}\right),
\end{eqnarray}
 where $\Delta f_{i}=\log Z_{i}$ and $\Delta f_{a}=\log Z_{a}$ are the
free energy shifts due to variable node ($i$) addition (and all its
function nodes) and function node ($a$)
addition~\cite{Zecchina-2008entropy} respectively. Actually $Z_{i}$
is the normalization constant of the full probability
$p_{i}^{J_{i}}$ and $Z_{a}$ the normalization constant of
$\hat{p}_{a}^{\boldsymbol{J}}$~\cite{MM-2009}.
$\hat{w}_{a}=\sum_{j\in\partial a}m_{j\rightarrow a}\xi_{j}^{a}$ and
$\hat{\sigma}_{a}=\sum_{j\in\partial a}(1-m_{j\rightarrow a}^{2})$.
Equations~(\ref{cavH01}) and~(\ref{cavH02}) can be solved by an
iterative procedure with a random initialization of the
corresponding messages. After the iteration converges, the entropy
landscape $s(d)$ from the fixed reference configuration
$\boldsymbol{J^{*}}$ can be computed according to the Legendre
transform Eqs.~(\ref{Legendre01}) and~(\ref{Legendre02}). The
computational complexity is of the order $\mathcal {O}(N^{2})$ for
this densely connected graphical model. $f(x)$ computed based on
Eq.~(\ref{fRS01}) does not depend on the reference configuration
since the change of $\xi_{i}^{b}\rightarrow-\xi_{i}^{b}$ does not
affect the final result, consistent with Eq.~(\ref{repf03}).

\begin{figure}
\centering
    \includegraphics[bb=17 17 285 221,width=0.8\textwidth]{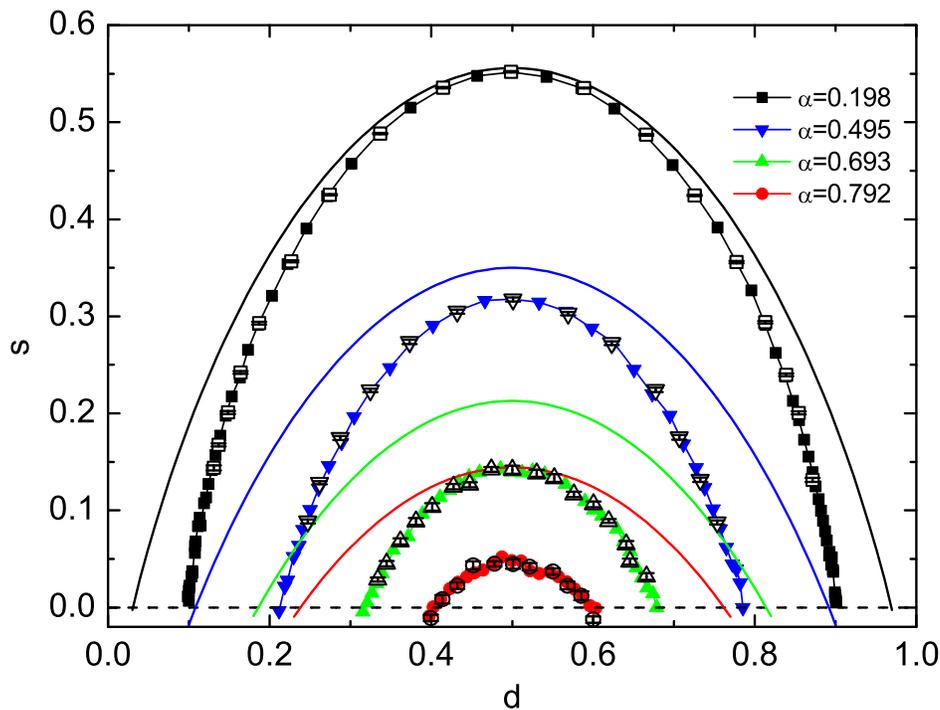}
  \caption{
  (Color online) Distance landscape from a reference
  configuration. The solid lines are the analytic annealed
  approximation (Eq.~(\ref{annS01})) for $\alpha=0.198,
  0.495,0.693,0.792$ (from the top to the bottom) respectively. The
  horizontal dashed line indicates the zero entropy value. The
  line connecting symbols is a guide to the eye. The empty symbols stay for the numerical simulation results on systems
  with $(N,P)=(1001,198),(1001,495),(1001,694),(1001,793)$ (from the top to the bottom) using message passing algorithms. The result is the average over $20$ random instances. Solid symbols are the
  replica symmetric results computed numerically by solving the
  saddle point equations.
  }\label{Sdr}
\end{figure}

\begin{figure}
\centering
    \includegraphics[bb=19 16 282 214,width=0.8\textwidth]{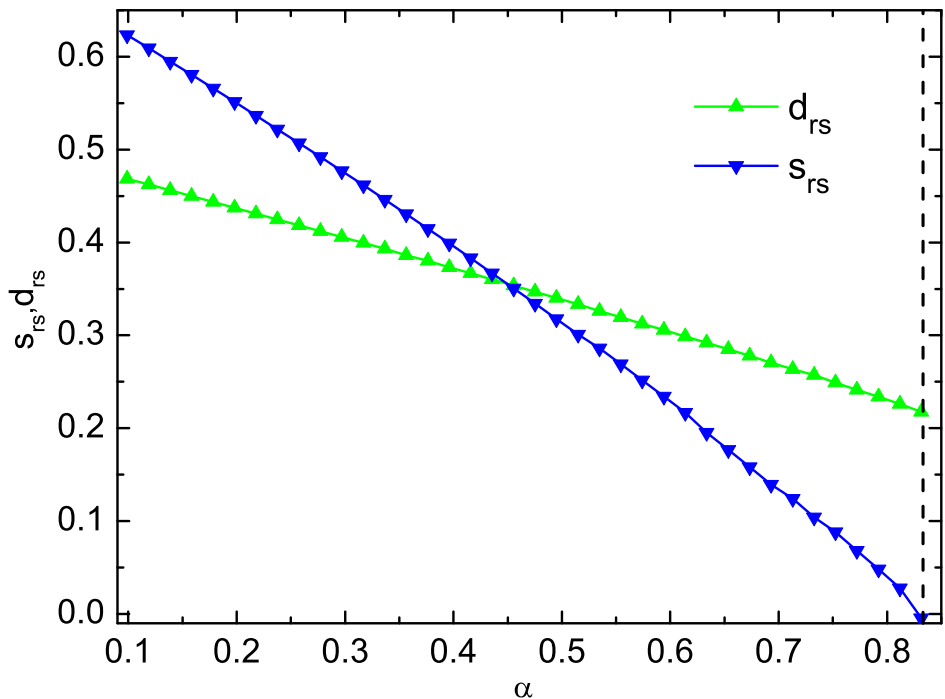}
  \caption{
  (Color online) Entropy density and typical distance between any
  two solutions as a function of constraint density. The vertical
  dashed
  line indicates the capacity for the binary perceptron.
  }\label{Sd}
\end{figure}

The distance landscape from a reference configuration is reported in
figure~\ref{Sdr}. We choose the reference configuration
$\boldsymbol{J^{*}}=\{J_{i}^{*}=1\}_{i=1}^{N}$ for simplicity. Other
choices of the reference configuration still yield the same behavior
of the landscape. Note that the annealed entropy provides an upper
bound for the RS one, and it roughly coincides with the RS one at
low $\alpha$ (around the maximal point) while the large deviation is
observed when $\alpha$ further increases. It is clear that most of
the solutions concentrate around the dominant point where the
maximum of the entropy is reached. When the given distance is larger
or less than certain values ($d>d_{{\rm max}}$ or $d<d_{{\rm
min}}$), the number of solutions at those distances becomes
exponentially small in $N$. In the intermediate range ($d\in[d_{{\rm
min}},d_{{\rm max}}]$), as the distance increases, the number of
solutions separated by the distance from the reference point in the
solution space increases first and then reaches the maximum which
dominates the typical value of the entropy in the original systems
(by setting $x=0$, see figure~\ref{Sd}). The maximum is then
followed by a decreasing trend as the distance is further increased.
This mean field behavior is confirmed by the numerical simulations
on large-size single random instances using the message passing
algorithms derived in Sec.~\ref{subsec:MPA01}. The consistency
between the mean field result obtained by replica approach and the
simulation result obtained on single random instances is clearly
shown in figure~\ref{Sdr}. The bell shape in figure~\ref{Sdr} is
similar to that observed in calculating the growth rate of expected
distance enumerator for a random code ensemble~\cite{Barg-2002}.
Note that as the constraint density increases, the distance range
where solutions exist shrinks, which illustrates clearly how the
solution space changes as more patterns are presented to the binary
perceptron.

We also compute the typical value of the entropy in the original
system (by setting $x=0$) and of the distance between any two
solutions as a function of the constraint density using the replica
method. The result is reported in figure~\ref{Sd}. Here we define
the typical value of the distance between any two solutions as
$d_{{\rm rs}}=\frac{1-q}{2}$ where $q$ is obtained from the
stationary value of Eq.~(\ref{repf03a}). The entropy vanishes at
$\alpha_{s}\simeq0.833$ with a finite typical value of
distance~\cite{Krauth-1989}. This typical distance is also in
accordance with that computed on single instances by sampling a
finite number of solutions~\cite{Huang-2010jstat}. Note that this
distance is evaluated here based on the RS ansatz. One can further
check its stability by the population dynamics~\cite{cavity-2001} on
the one-step replica symmetry breaking ($1$RSB)
solution~\cite{MM-2009} where we define two typical distances: one
is inter-cluster distance $d_{0}=\frac{1-\left[\left<\tanh
h\right>^{2}\right]}{2}$ where $\left<\cdot\right>$ means the
average over clusters and $\left[\cdot\right]$ over the disorder,
and the other is intra-cluster distance defined by
$d_{1}=\frac{1-\left[\left<\tanh^{2}h\right>\right]}{2}$~\cite{Biro-2000}
where $h$ is the local field defined in Eq.~(\ref{cavH01}) by
including all contributions of patterns around the synaptic weight
($x=0$). In general, solutions within a single cluster are separated
by a sub-extensive number of synaptic weights while any two clusters
are separated by an extensive number of synaptic weights. Our
numerical simulations confirmed that $d_{0}$ and $d_{1}$ will turn
out to be identical (equal to $d_{{\rm rs}}$) after sufficient
iterations implying that the RS ansatz is unbroken below the
capacity. However, for constraint density close to the capacity, one
needs a much larger sampling interval (by way of the Metropolis
importance sampling method)~\cite{ZH-2008} in the population
dynamics algorithm. To probe the fine structure of the connection
pattern in the solution space, we study the distance landscape of
solution-pairs in the following section. This is similar to the
study of the spherical $p$-spin model in the presence of an
attractive coupling with quenched configuration in
Ref.~\cite{Franz-1997prl}, however, the rich information about the
solution space structure of the current problem can also be attained
by calculating the distance landscape of solution-pairs.

\section{Distance landscape of solution-pairs} \label{sec:sdp}

The geometrical property of the solution space can also be studied
by counting the number of solution-pairs with a predefined distance
$d$, equivalently an overlap of value $\tilde{q}$. Actually this entropy
value may be much larger than the entropy density of the original
problem (which is obtained by setting $x=0$ in Eq.~(\ref{fRS01})).
As we shall present later, this case becomes more involved for the
binary perceptron with an increasing computational cost.

Considering distance between solutions, we write the partition
function as
\begin{equation}\label{PTF03}
Z=\sum_{\boldsymbol{J}^{1},\boldsymbol{J}^{2}}\prod_{\mu}\Theta\left(\frac{1}{\sqrt{N}}\sum_{i}J_{i}^{1}\xi_{i}^{\mu}\right)
\Theta\left(\frac{1}{\sqrt{N}}\sum_{i}J_{i}^{2}\xi_{i}^{\mu}\right)\exp\left[x\sum_{i}J_{i}^{1}J_{i}^{2}\right]
\end{equation}
where the coupling field $x$ is used to control the distance between
a pair of solutions $(\boldsymbol{J}^{1},\boldsymbol{J}^{2})$ and
the associated overlap
$\tilde{q}\equiv\frac{1}{N}\sum_{i}J_{i}^{1}J_{i}^{2}$. This partition
function has been used to predict optimal coupling field for a
multiple random walking strategy to find a solution for the
perceptronal learning problem~\cite{Huang-2011epl}. In the following
sections, we present an annealed computation as well as RS
computation of the distance landscape $s(d)$. Note that in this
setting, Eqs.~(\ref{PTF02}),~(\ref{Legendre01})
and~(\ref{Legendre02}) can also be used but here $d$ should be
understood as the distance separating two solutions in the weight
space.

\subsection{Annealed approximation for $s(d)$}
\label{subsec:ann02}

Following the same techniques used in Sec.~\ref{subsec:ann01}, we
obtain the annealed free energy density as (see also
Ref.~\cite{Huang-2011epl})
\begin{equation}\label{annf02}
f_{\rm{ann}}=\max_{\tilde{q},\hat{q}}\left\{-\hat{q}\tilde{q}+x\tilde{q}+\log(4\cosh\hat{q})+\alpha\log\int_{0}^{\infty}DtH\left(-\frac{\tilde{q}t}{\sqrt{1-\tilde{q}^{2}}}\right)\right\}.
\end{equation}
The maximization with respect to $\tilde{q}$ and $\hat{q}$ leads to the
following saddle point equation
\begin{eqnarray}\label{sde01}
\tilde{q}&=&\tanh\hat{q},\\
\hat{q}&=&x+\frac{\alpha}{\sqrt{1-\tilde{q}^{2}}{\rm
arccot}\left(-\frac{\tilde{q}}{\sqrt{1-\tilde{q}^{2}}}\right)},\label{sde02}
\end{eqnarray}
where the identity
$\int_{0}^{\infty}DtH\left(-\frac{\tilde{q}t}{\sqrt{1-\tilde{q}^{2}}}\right)=\frac{1}{2\pi}{\rm
arccot}\left(-\frac{\tilde{q}}{\sqrt{1-\tilde{q}^{2}}}\right)$ has been used. Using
Eq.~(\ref{Legendre01}) and the above saddle point equation, we get
the final expression for $s_{{\rm ann}}(d)$:
\begin{eqnarray}
s_{\rm{ann}}(d)&=&\log2-(1-d)\log(1-d)-d\log
d\nonumber\\
&+&\alpha\log\left[\frac{1}{2\pi}{\rm
arccot}\left(-\frac{1-2d}{2\sqrt{d(1-d)}}\right)\right],\label{anns02}
\end{eqnarray}
where $s_{{\rm ann}}(0)=(1-\alpha)\log2$ which is actually the
annealed entropy density of the original system~\cite{Krauth-1989}.
If $\alpha=0$, then $s_{{\rm ann}}(0)=\log2$ which is in accord with
the fact that the number of solution-pairs with a distance $d=0$
should be the total number of solutions $2^{N}$ if no constraints
are present.

\subsection{Replica symmetric computation of $s(d)$}
\label{subsec:RS02}

In this section, we derive the free energy density $f(x)$ for the
landscape of solution-pairs under the replica symmetric
approximation, using the replica trick introduced in
Sec.~\ref{subsec:RS01}. Since the partition function in this case
involves a sum of all possible configurations of two synaptic weight
vectors, the computation becomes a bit complicated. The disorder
average of the integer power of the partition function can be
evaluated as
\begin{eqnarray}\label{repair01}
\fl\left<Z^{n}\right>=\sum_{\{\boldsymbol{J}^{1,a},\boldsymbol{J}^{2,a}\}}\prod_{a<b}\int
dq_{1}^{ab}dq_{2}^{ab}dr^{ab}\delta\left(q_{1}^{ab}-\frac{1}{N}\sum_{i}J_{i}^{1,a}J_{i}^{1,b}\right)
\delta\left(q_{2}^{ab}-\frac{1}{N}\sum_{i}J_{i}^{2,a}J_{i}^{2,b}\right)\nonumber\\
\times\delta\left(r^{ab}-\frac{1}{N}\sum_{i}J_{i}^{1,a}J_{i}^{2,b}\right)\prod_{a}\int
dR^{aa}\delta\left(R^{aa}-\frac{1}{N}\sum_{i}J_{i}^{1,a}J_{i}^{2,a}\right)\nonumber\\
\times\left<\prod_{\mu,a}\Theta(w_{1}^{\mu,a})\Theta(w_{2}^{\mu,a})\right>e^{x\sum_{a,i}J_{i}^{1,a}J_{i}^{2,a}},
\end{eqnarray}
where
$w_{1}^{\mu,a}\equiv\frac{1}{\sqrt{N}}\sum_{i}J_{i}^{1,a}\xi_{i}^{\mu}$
and
$w_{2}^{\mu,a}\equiv\frac{1}{\sqrt{N}}\sum_{i}J_{i}^{2,a}\xi_{i}^{\mu}$.
Under the replica symmetric ansatz, the disorder average is carried
out as
\begin{equation}
\left<\prod_{a}\Theta(w_{1}^{\mu,a})\Theta(w_{2}^{\mu,a})\right>_{\boldsymbol{\xi}^{\mu}}=\int
D\boldsymbol{z}\left[\int Dy H(y_{1})H(y_{2})\right]^{n},
\end{equation}
where $\int D\boldsymbol{z}\equiv\int Dz_{1}\int Dz_{2}\int Dt$,
$y_{x}=-\frac{\sqrt{R-r}y+\sqrt{q-r}z_{x}+\sqrt{r}t}{\sqrt{1-q-R+r}}(x=1,2)$
and we have used $q_{1}^{ab}=q_{2}^{ab}=q,r^{ab}=r,R^{aa}=R$ under
the RS ansatz. After the computation of the summation in
Eq.~(\ref{repair01}) by using
$\sum_{a,b}J_{i}^{a}J_{i}^{b}=\frac{\left(\sum_{a}J_{i}^{1,a}+\sum_{b}J_{i}^{2,b}\right)^{2}-\left(\sum_{a}J_{i}^{1,a}\right)^{2}-\left(\sum_{b}J_{i}^{2,b}\right)^{2}}{2}$
and the Hubbard-stratonovich transform, we get the replica symmetric
free energy density
\begin{eqnarray}\label{repair02}
\fl f(x)=\alpha\int D\boldsymbol{z}\log
F_{1}(q,R,r)+xR+\hat{q}(q-1)+\frac{1}{2}r\hat{r}-R\hat{R}+\int
D\boldsymbol{\hat{z}}\log F_{2}(\hat{q},\hat{R},\hat{r}),
\end{eqnarray}
where $\int D\boldsymbol{\hat{z}}\equiv\int D\hat{z}_{1}\int
D\hat{z}_{2}\int D\hat{z}_{3}$, $F_{1}(q,R,r)=\int Dy
H(y_{1})H(y_{2})$ and
$F_{2}(\hat{q},\hat{R},\hat{r})=2e^{a_{3}}\cosh(a_{1}+a_{2})+2e^{-a_{3}}\cosh(a_{1}-a_{2})$
where
$a_{x}=\sqrt{\hat{q}-\hat{r}/2}\hat{z}_{x}+\sqrt{\hat{r}/2}\hat{z}_{3}(x=1,2),a_{3}=\hat{R}-\hat{r}/2$.
The RS order parameters $(q,R,r,\hat{q},\hat{R},\hat{r})$ are
determined by the following self-consistent equations
\begin{eqnarray}\label{pairsd}
\fl\hat{R}&=x+\frac{\alpha}{1-q-R+r}\int D\boldsymbol{z}\frac{\int Dy G(y_{1})G(y_{2})}{\int Dy H(y_{1})H(y_{2})},\label{paira}\\
\fl\hat{r}&=\frac{2\alpha}{1-q-R+r}\int D\boldsymbol{z}\frac{\int Dy
G(y_{1})H(y_{2})\int Dy G(y_{2})H(y_{1})}{\left[\int
Dy H(y_{1})H(y_{2})\right]^{2}},\\
\fl\hat{q}&=\frac{\hat{r}}{2}+\frac{\alpha}{2(1-q-R+r)}\int
D\boldsymbol{z}\left[\frac{\int Dy\left[G(y_{1})H(y_{2})-
G(y_{2})H(y_{1})\right]}{\int Dy H(y_{1})H(y_{2})}\right]^{2},\label{pairb}\\
 \fl R&=\int D\boldsymbol{\hat{z}}\frac{\tanh a_{3}+\tanh a_{1}\tanh
a_{2}}{1+\tanh a_{3}\tanh a_{1}\tanh
a_{2}},\label{pairR}\\
 \fl r&=\int D\boldsymbol{\hat{z}}\frac{\tanh
a_{3}(\tanh^{2}a_{1}+\tanh^{2}a_{2})+\tanh a_{1}\tanh
a_{2}(1+\tanh^{2}a_{3})}{(1+\tanh a_{3}\tanh a_{1}\tanh
a_{2})^{2}},\\
\fl q&=r+\int D\boldsymbol{\hat{z}}\frac{(\tanh a_{3}-1)^{2}(\tanh
a_{1}-\tanh a_{2})^{2}}{2(1+\tanh a_{3}\tanh a_{1}\tanh a_{2})^{2}}.
\end{eqnarray}
In the derivation of the above saddle point equations, we have used
a useful property of the Gaussian measure $\int Dz zF(z)=\int
DzF'(z)$ where $F'(z)$ is the derivative of the function $F(z)$ with
respect to $z$. After the fixed point of the above saddle point
equations is obtained, one can compute the entropy density
$s=f(x)-xR$ with $d=\frac{1-R}{2}$. Note that $R-r$ may become
negative, in this case we replace $R$ and $r$ by $-R$ and $-r$
respectively, $y$ by $-y$ only for $y_{2}$ in Eqs.~(\ref{paira})
to~(\ref{pairb}).

\subsection{Message passing equations for single instances}
\label{subsec:MPA02}

By analogy with definitions in Sec.~\ref{subsec:MPA01}, we define by
$p_{i\rightarrow a}^{J_{i}^{1},J_{i}^{2}}$ the probability that the
synaptic weight $i$ takes a two-component vector state
$(J_{i}^{1},J_{i}^{2})$ in the absence of constraint $a$ and by
$\hat{p}_{b\rightarrow i}^{J_{i}^{1},J_{i}^{2}}$ the probability
that constraint $b$ is satisfied given the vector state
$(J_{i}^{1},J_{i}^{2})$ of weight $i$. These two cavity
probabilities obey the following recursive equations
\begin{eqnarray}\label{bp2}
\fl p_{i\rightarrow
a}^{J_{i}^{1},J_{i}^{2}}&=&\frac{1}{Z_{i\rightarrow
a}}e^{xJ_{i}^{1}J_{i}^{2}}\prod_{b\in\partial i\backslash
a}\hat{p}_{b\rightarrow i}^{J_{i}^{1},J_{i}^{2}},\\
\fl\hat{p}_{b\rightarrow
i}^{J_{i}^{1},J_{i}^{2}}&=&\sum_{\{J_{j}^{1},J_{j}^{2},j\in\partial
b\backslash
i\}}\Theta\left(\frac{1}{\sqrt{N}}\sum_{j}J_{j}^{1}\xi_{j}^{b}\right)\Theta\left(\frac{1}{\sqrt{N}}\sum_{j}J_{j}^{2}\xi_{j}^{b}\right)
\prod_{j\in\partial b\backslash i}p_{j\rightarrow
b}^{J_{j}^{1},J_{j}^{2}},\label{bp3}
\end{eqnarray}
where $Z_{i\rightarrow a}$ is a normalization constant. In fact the
belief propagation equations~(\ref{bp2}) and~(\ref{bp3}) correspond
to the stationary point of the Bethe free energy function of the
current system~\cite{cavity-2003,Yedidia-2005}. The exchange of
$\boldsymbol{J}^{1}$ and $\boldsymbol{J}^{2}$ does not change the
partition function Eq.~(\ref{PTF03}), thus the cavity probabilities
have the property that $p_{i\rightarrow a}^{+1,-1}=p_{i\rightarrow
a}^{-1,+1}$ and $\hat{p}_{b\rightarrow
i}^{+1,-1}=\hat{p}_{b\rightarrow i}^{-1,+1}$. This symmetry property
will simplify the following analysis a lot.

To simplify Eqs.~(\ref{bp2}) and~(\ref{bp3}), we need the joint
distribution of $w_{b\rightarrow i}^{1}$ and $w_{b\rightarrow
i}^{2}$ where $w_{b\rightarrow
i}^{1}\equiv\frac{1}{\sqrt{N}}\sum_{j\neq i}J_{j}^{1}\xi_{j}^{b}$
and $w_{b\rightarrow i}^{2}\equiv\frac{1}{\sqrt{N}}\sum_{j\neq
i}J_{j}^{2}\xi_{j}^{b}$. Since we impose a distance constraint upon
two solutions $\boldsymbol{J}^{1}$ and $\boldsymbol{J}^{2}$ in
Eq.~(\ref{PTF03}), there exists correlation between these two
normally distributed random numbers and this correlation is
characterized by the correlation coefficient
\begin{equation}\label{rho}
\hat{\rho}_{b\rightarrow i}=\frac{\sum_{j\in\partial b\backslash
i}(q_{j\rightarrow b}-m_{j\rightarrow
b}^{2})}{\hat{\sigma}_{b\rightarrow i}}
\end{equation}
due to the symmetry property. Based on Eq.~(\ref{bp2}), messages $q_{j\rightarrow b}$ and
$m_{j\rightarrow b}$ are determined respectively by the following
equations,
\begin{eqnarray}
q_{j\rightarrow b}&=&\frac{e^{x}\left[\prod_{a\in\partial
j\backslash b}\hat{p}_{a\rightarrow j}^{+1,+1}+\prod_{a\in\partial
j\backslash b}\hat{p}_{a\rightarrow
j}^{-1,-1}\right]-2e^{-x}\prod_{a\in\partial j\backslash
b}\hat{p}_{a\rightarrow j}^{+1,-1}}{e^{x}\left[\prod_{a\in\partial
j\backslash b}\hat{p}_{a\rightarrow j}^{+1,+1}+\prod_{a\in\partial
j\backslash b}\hat{p}_{a\rightarrow
j}^{-1,-1}\right]+2e^{-x}\prod_{a\in\partial j\backslash
b}\hat{p}_{a\rightarrow
j}^{+1,-1}},\label{q}\\
 m_{j\rightarrow
b}&=&\frac{e^{x}\left[\prod_{a\in\partial j\backslash
b}\hat{p}_{a\rightarrow j}^{+1,+1}-\prod_{a\in\partial j\backslash
b}\hat{p}_{a\rightarrow
j}^{-1,-1}\right]}{e^{x}\left[\prod_{a\in\partial j\backslash
b}\hat{p}_{a\rightarrow j}^{+1,+1}+\prod_{a\in\partial j\backslash
b}\hat{p}_{a\rightarrow j}^{-1,-1}\right]+2e^{-x}\prod_{a\in\partial
j\backslash b}\hat{p}_{a\rightarrow j}^{+1,-1}}.\label{cavH03}
\end{eqnarray}
Therefore, both $w_{b\rightarrow i}^{1}$ and $w_{b\rightarrow
i}^{2}$ obey a bivariate normal distribution and
$\hat{p}_{b\rightarrow i}^{J_{i}^{1},J_{i}^{2}}$ is reduced to be
\begin{equation}\label{Phat02}
\hat{p}_{b\rightarrow
i}^{J_{i}^{1},J_{i}^{2}}=\int_{-\frac{J_{i}^{2}\xi_{i}^{b}+\hat{w}_{b\rightarrow
i}}{\sqrt{\hat{\sigma}_{b\rightarrow
i}}}}^{\infty}DtH\left(-\frac{J_{i}^{1}\xi_{i}^{b}+\hat{w}_{b\rightarrow
i}}{\sqrt{(1-\hat{\rho}_{b\rightarrow
i}^{2})\hat{\sigma}_{b\rightarrow
i}}}-\frac{\hat{\rho}_{b\rightarrow
i}t}{\sqrt{1-\hat{\rho}_{b\rightarrow i}^{2}}}\right)
\end{equation}
where $\hat{w}_{b\rightarrow i}=\sum_{j\in\partial b\backslash
i}m_{j\rightarrow b}\xi_{j}^{b}$ and $\hat{\sigma}_{b\rightarrow
i}=\sum_{j\in\partial b\backslash i}(1-m_{j\rightarrow b}^{2})$. The
overlap $\tilde{q}$ is determined by $\tilde{q}(x)=\frac{1}{N}\sum_{i}\tilde{q}_{i}$ where
$\tilde{q}_{i}$ is given by
\begin{equation}
\tilde{q}_{i}=\frac{e^{x}\left[\prod_{b\in\partial i}\hat{p}_{b\rightarrow
i}^{+1,+1}+\prod_{b\in\partial i}\hat{p}_{b\rightarrow
i}^{-1,-1}\right]-2e^{-x}\prod_{b\in\partial i}\hat{p}_{b\rightarrow
i}^{+1,-1}}{e^{x}\left[\prod_{b\in\partial i}\hat{p}_{b\rightarrow
i}^{+1,+1}+\prod_{b\in\partial i}\hat{p}_{b\rightarrow
i}^{-1,-1}\right]+2e^{-x}\prod_{b\in\partial i}\hat{p}_{b\rightarrow
i}^{+1,-1}}.
\end{equation}
 Eq.~(\ref{Phat02}) is more computationally demanding
than Eq.~(\ref{Phat}) since an additional numerical integral is
required to compute $\hat{p}$ here. However, the integral in
Eq.~(\ref{Phat02}) can be approximated by
$c_{0}H\left(\frac{c+c_{1}}{\sqrt{c_{2}}}\right)$ if we write the
right hand side of Eq.~(\ref{Phat02}) as $\int_{c}^{\infty}Dte^{\log
H\left(a-bt\right)}$ and expand $H\left(a-bt\right)$ up to the
second order in $bt$. The constants $c_{0}$, $c_{1}$ and $c_{2}$ can
be determined as a function of $a$ and $b$. Therefore, this
approximation is accurate only when large $bt$ has vanishing
contribution to the integral.

The free energy shift due to variable node addition (and all its
adjacent constraints) can be obtained as $\Delta f_{i}=\log Z_{i}$
and the free energy shift due to constraint addition $\Delta
f_{a}=\log Z_{a}$ where
\begin{eqnarray}\label{fRS02}
Z_{i}&=&e^{x}\left[\prod_{b\in\partial i}\hat{p}_{b\rightarrow
i}^{+1,+1}+\prod_{b\in\partial i}\hat{p}_{b\rightarrow
i}^{-1,-1}\right]+2e^{-x}\prod_{b\in\partial i}\hat{p}_{b\rightarrow
i}^{+1,-1},\\
Z_{a}&=&\int_{-\frac{\hat{w}_{a}}{\sqrt{\hat{\sigma}_{a}}}}^{\infty}DtH\left(-\frac{\hat{w}_{a}}{\sqrt{(1-\hat{\rho}_{a}^{2})\hat{\sigma}_{a}}}-\frac{\hat{\rho}_{a}t}{\sqrt{1-\hat{\rho}_{a}^{2}}}\right),
\end{eqnarray}
where $\hat{w}_{a}=\sum_{j\in\partial a}m_{j\rightarrow
a}\xi_{j}^{a}$, $\hat{\sigma}_{a}=\sum_{j\in\partial
a}(1-m_{j\rightarrow a}^{2})$ and
$\hat{\rho}_{a}=\frac{\sum_{i\in\partial a}(q_{i\rightarrow
a}-m_{i\rightarrow a}^{2})}{\hat{\sigma}_{a}}$. The free energy
density can then be obtained using Eq.~(\ref{fRS01}) and the entropy
landscape can be obtained correspondingly. The recursive equations
Eqs.~(\ref{rho}),~(\ref{q}),~(\ref{cavH03}) and~(\ref{Phat02}) can
be solved by an iterative procedure similar to that used in
Sec.~\ref{subsec:MPA01}.

\begin{figure}
\centering
    \includegraphics[bb=19 10 286 218,width=0.8\textwidth]{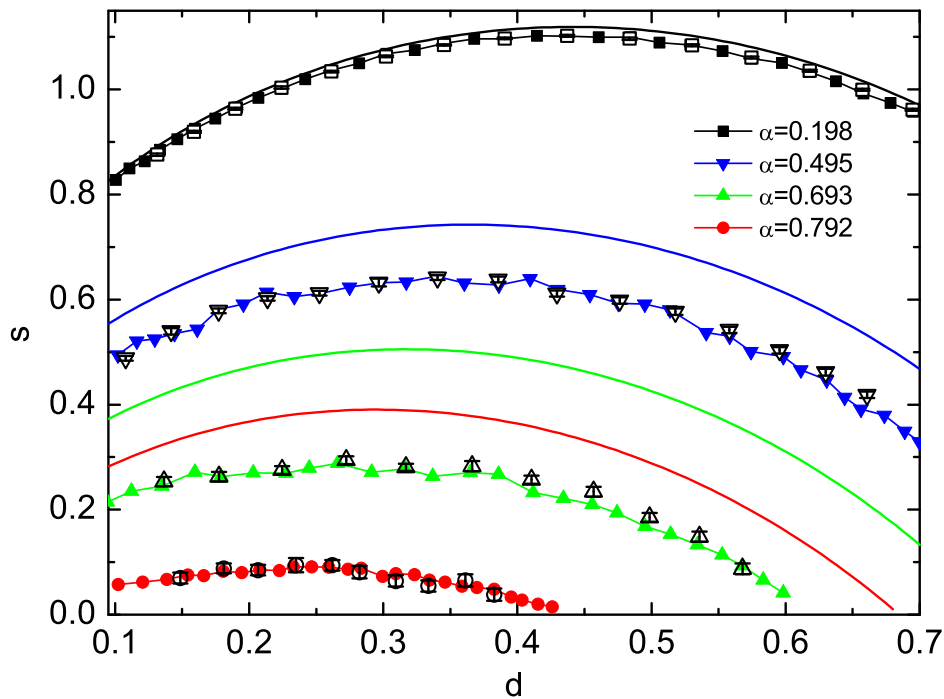}
  \caption{
  (Color online) Distance landscape of solution-pairs with a predefined distance $d$. The solid lines are the analytic annealed
  approximation (Eq.~(\ref{anns02})) for $\alpha=0.198,
  0.495,0.693,0.792$ (from the top to the bottom) respectively. The
  line connecting symbols is a guide to the eye. The empty symbols stay for the numerical simulation results on systems
  with $(N,P)=(501,99),(501,248),(501,347),(501,397)$ (from the top to the bottom) using message passing algorithms. The result is the average over $20$ random instances. Solid symbols are the
  replica symmetric results computed numerically by solving the
  saddle point equations.
  }\label{Sdp}
\end{figure}

\begin{figure}
\centering
    \includegraphics[bb=17 15 291 218,width=0.8\textwidth]{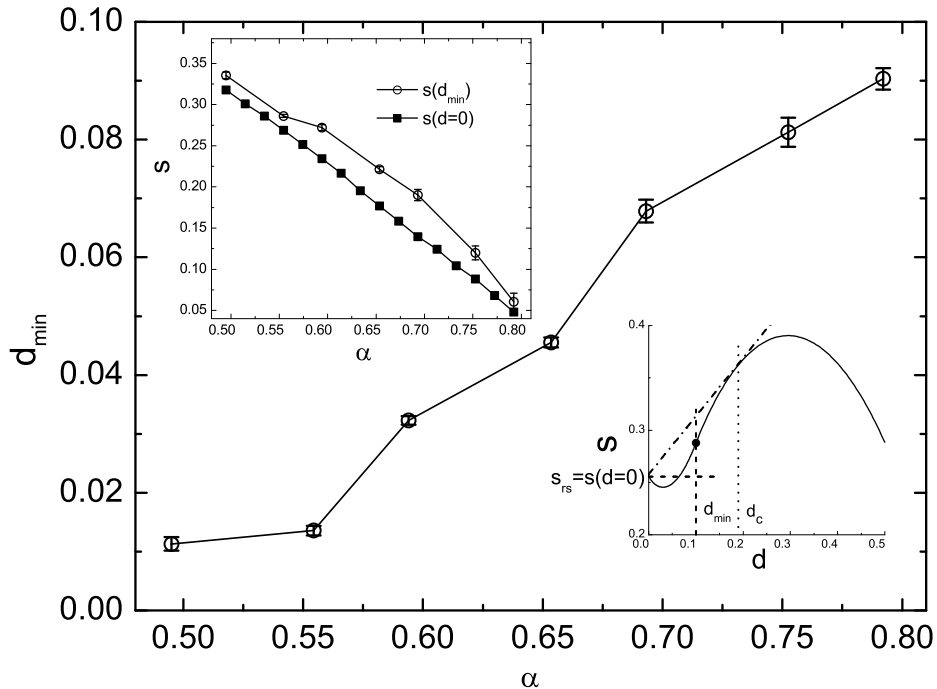}
  \caption{
   Minimal distance $d_{{\rm min}}$ as
   a function of the constraint density. The data points are computed by solving numerically the saddle point equations in Sec.~\ref{subsec:RS02}. The error bars characterize the numerical fluctuations
   from ten
   different random initializations. The upper inset shows the
   corresponding entropy values. The lower inset shows {\em
   schematically}
   the typical concave and non-concave behavior of the entropy
   landscape, where the horizontal dashed line indicates $s_{{\rm
   rs}}$ and the vertical dashed line denotes $d_{{\rm min}}$, and
   the black point $(d_{{\rm min}},s(d_{{\rm min}}))$ marks the first change of the concavity, i.e., $\frac{\partial^{2}s(d)}{\partial d^{2}}=0$.
   The vertical dotted line marks the first order thermodynamic transition point
   $d_{c}$, where the dash-dotted line going through $(0,s_{{\rm rs}})$ touches the concave part of
   $s(d)$ and has the slope $2x_{c}$. Note that $d_{{\rm min}}$
   corresponds to the spinodal point $x_{s}$ ($\geq x_{c}$), i.e., $\frac{\partial s(d)}{\partial d}|_{d=d_{{\rm min}}}=2x_{s}$.
  }\label{gap}
\end{figure}

\begin{figure}
\centering
          \includegraphics[bb=16 15 277 216,width=0.8\textwidth]{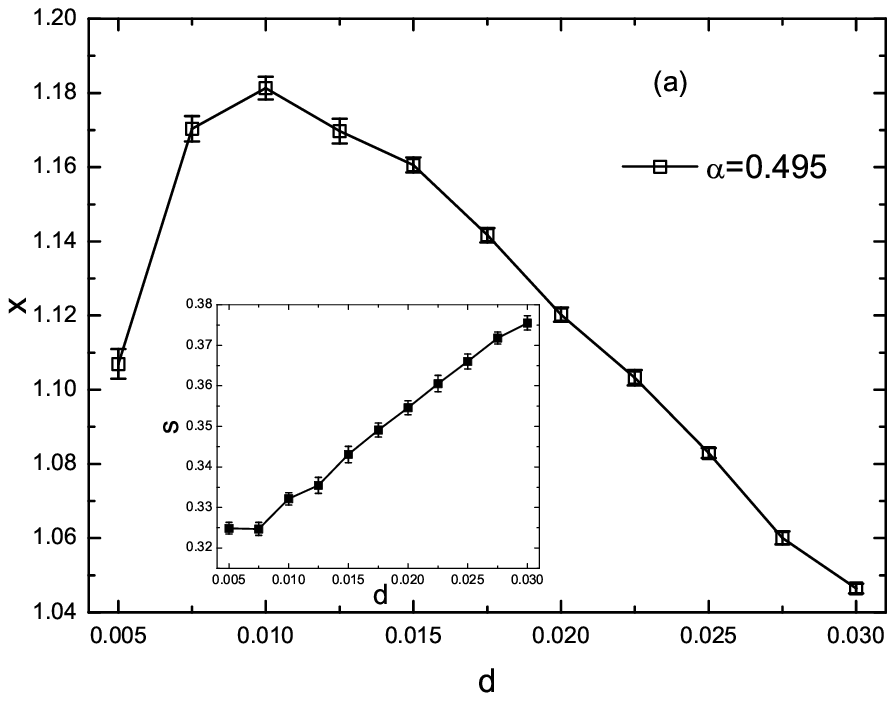}
     \vskip .05cm
     \includegraphics[bb=28 18 289 217,width=0.8\textwidth]{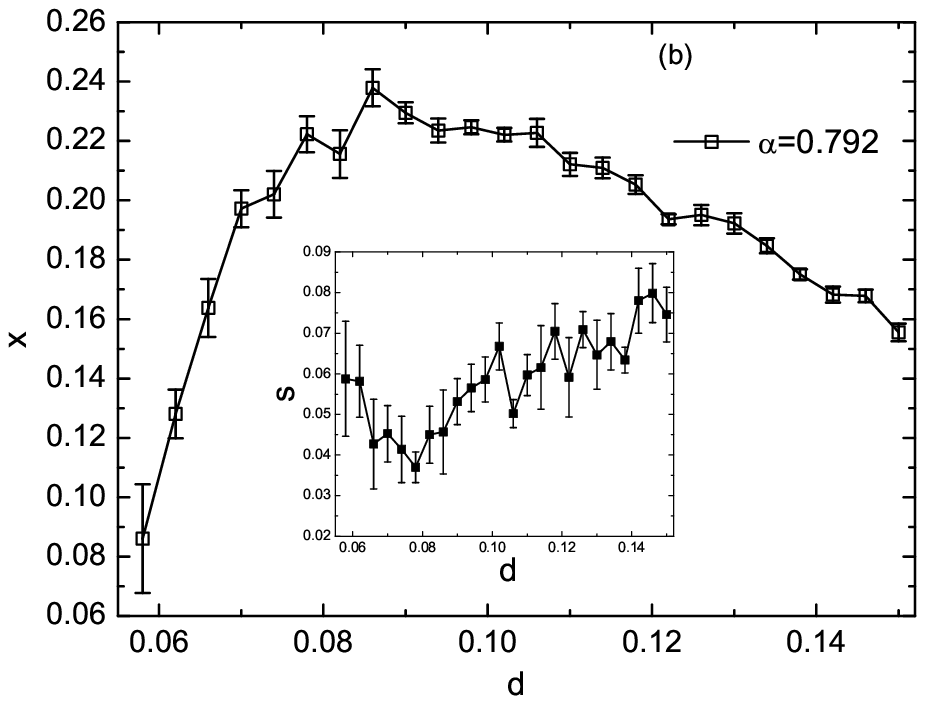}\vskip .05cm
  \caption{
     Compatible coupling field $x$ for fixed $d$. The data points are computed by solving numerically the saddle point equations in Sec.~\ref{subsec:RS02}. The error bars characterize the numerical fluctuations from
  ten different random initializations. Inset: the
     corresponding entropy curve as a function of $d$.
      (a) $\alpha=0.495$. (b) $\alpha=0.792$.
   }\label{nonconcav}
 \end{figure}

 As is seen from figure~\ref{Sdp}, the entropy
density increases smoothly until a maximum is reached for
$\alpha=0.198$ and then decreases as the distance further grows.
Interestingly, this behavior observed in figure~\ref{Sdp} can be
well fitted by the annealed approximation keeping the concavity of
the entropy function. However, as $\alpha$ increases, large
deviation from the annealed approximation occurs. The mean field
calculations are supported by the numerical simulations on single
instances using the proposed message passing algorithms, as shown in
figure~\ref{Sdp}. The distance corresponding to the maximum of the
entropy landscape curve in figure~\ref{Sdp} is actually the typical
distance $d_{{\rm rs}}$ calculated in figure~\ref{Sd}, and $s(d=0)$
recovers the typical entropy density of the original problem. By
taking the limit $R\rightarrow 1$ in Eq.~(\ref{repair02}), one can
show that $s(d=0)=f(x=0)$ where $f(x)$ is given by
Eq.~(\ref{repf03}). As the constraint density increases, the maximal
point of the entropy curve moves to the left, however,
solution-pairs still maintain a relatively broad distribution in the
solution space when $\alpha$ approaches $\alpha_{s}$, e.g., $d_{\rm
max}={\rm argmax}_{d}\{s(d)=0\}\simeq0.332$ at $\alpha=0.82$
($d_{{\rm rs}}\simeq0.222$), which may be responsible for the
algorithmic hardness in this region.

As $\alpha$ increases, the message passing algorithm requires a
large number of iteration steps to converge (especially at small
distances) and additionally a computationally expensive Monte Carlo
integral involved in Eq.~(\ref{Phat02}) cannot be avoided. On the
other hand, when $\alpha$ is large enough, one can easily observe a
rapid growth of the order parameter $R$ to unity, i.e., at some
critical coupling field $x_{c}$, $R$ changes sharply from a value
smaller than one to one. This implies that at $x_{c}$, $R=1$ becomes
a globally stable solution of the saddle point equations in
Sec.~\ref{subsec:RS02}. The first order thermodynamic transition is
signalled by the change of the concavity at $d=d_{{\rm min}}>0$. We
define $d_{{\rm min}}$ as the minimal distance before $R=1$ becomes
a unique stable solution. Figure~\ref{gap} shows the entropy gap and
$d_{{\rm min}}$ as a function of the constraint density. The
corresponding coupling field $x_{s}$ marks the point where the
concavity starts to change, i.e., $\frac{\partial^{2}s(d)}{\partial
d^{2}}=0$, as shown in the lower inset of figure~\ref{gap}
($x_{s}\geq x_{c}$). After $x_{s}$, $R=1$ becomes the unique stable
solution of the saddle point equations. Note that in the entropy
gap, there exists a non-concave part of the entropy curve (for small
distances), which can only be obtained by fixing $d$ instead of $x$
and searching for a compatible $x$ (by the secant method). The
result is shown in figure~\ref{nonconcav} for $\alpha=0.495$ and
$0.792$. The compatible $x$ for small distances (the left branch) is
smaller than $x_{s}$. When $x>x_{c}$, the right branch is no longer
globally stable solution but becomes metastable solution of the
saddle point equations until $x=x_{s}$, i.e., the spinodal point is
reached. By fixing $x_{c}\leq x\leq x_{s}$, one typically observes
the right branch or $R=1$, which describes the equilibrium
properties of the Boltzmann measure in Eq.~(\ref{PTF03}). Thus, the
non-concave behavior observed in $d\in(0,d_{{\rm min}})$ is
thermodynamically non-dominant and unstable, suggesting that the
solution space is made of isolated solutions instead of separated
clusters of exponentially many close-by solutions, and this behavior
becomes much more evident as $\alpha$ increases. This explains why
the multiple random walking strategy is extremely difficult to find
a solution by tuning the coupling field at high $\alpha$ and large
$N$~\cite{Huang-2011epl}.

We argue that for any finite $\alpha$, the slope of the entropy
curve $s(d)$ near to $d=0$ ($R=1$) tends to negative infinity.
Letting $\epsilon=1-R\rightarrow 0$, we can obtain $\frac{{\rm
d}S(R)}{{\rm d}R}=\alpha
C_{p}\epsilon^{-1/2}+C+\frac{1}{2}\ln\epsilon$, where $C_{p}$ is a
positive constant independent of $\alpha$ and $\epsilon$, and $C$ is
a constant as well. The derivation is given in~\ref{sec:appendix-a}.
Thus, as long as $\alpha>0$, the non-concave part exists in the
entropy curve for the interval $(0,d_{{\rm min}})$, implying that
such solution-pairs are exponentially less numerous than the typical
ones. Furthermore, $R=1$ is always a stable solution, and in this
case $s(R=1)=s_{{rs}}$.

As shown in figure~\ref{gap}$, d_{{\rm min}}$ increases as $\alpha$
grows, making a uniform sampling of solutions extremely hard. In
addition, $d_{{\rm min}}$ seems to grow continuously, being the
order of $\mathcal {O}(10^{-3})$ or less for $\alpha<0.5$. The
isolations of solutions can be explained by the nature of the hard
constraints~\cite{Frozen-2004prl} for the binary perceptron. Unlike
the random $K$-SAT and graph-coloring
problems~\cite{Zhou-2010epjb,Lenka-2008Jphys}, the hard constraint
in the binary perceptron problem implies that the synaptic weight on
one node in the factor graph is completely determined by the values
of other nodes. But for finite $N$, solutions may not be strictly
isolated. This explains why some local search heuristics can find a
solution when $N$ and $\alpha$ is not large enough. As $\alpha$
increases, some frozen solutions are more prone to disappear, thus
solutions become much more far apart from each other, as shown by
increasing $d_{{\rm min}}$ in figure~\ref{gap}. However, the
thermodynamic properties can still be derived from the RS solution
before $\alpha_{s}$. We conjecture that clustering and freezing
coexist for $\alpha<\alpha_{s}$, which is consistent with the
computation in Ref.~\cite{Kaba-09} in the sense that the total
entropy (displayed in figure~\ref{Sd}) $s_{{\rm rs}}=\Sigma(s)+s$
where $\Sigma(s)$ is the complexity of clusters of entropy density
$s$ and $s=0$ for the current problem. The structure of the solution
space is described by the dynamical one-step replica symmetry
breaking scenario (at the Parisi parameter
$m=1$)~\cite{Montanari-2008jstat}. By contrast, in the random
$K$-XORSAT, there exists a phase where going from a solution to
another one involves a large but finite Hamming
distance~\cite{Mezard-2003jstatphys,Mora-2006jstat} and in locked
constraint satisfaction problem, there exists logarithmic Hamming
distance separation between solutions in the liquid
phase~\cite{Lenka-08}. For the binary perceptron problem, we can say
that the solution space is simple in the sense that it is made of
isolated solutions instead of separated clusters of exponentially
many solutions; however, it becomes rather difficult to find a
solution via stochastic local search algorithm. Below $\alpha_{s}$,
$s_{{\rm rs}}>0$, meaning that there exist exponentially many
solutions, but they are widely dispersed (much more apparent at
large $\alpha$). In other words, solution-pairs maintain a
relatively broad distribution.

\section{Discussion and Conclusion}
\label{sec_con}

The typical property of the distance landscape either from a
reference configuration or for pairs of solutions is studied. For
the first distance landscape, as the distance increases, the number
of associated solutions grows first and then reaches its maximum
(dominating the typical value of the entropy in the original system)
followed by a decreasing behavior. This typical trend is confirmed
by the numerical simulations on single instances using the proposed
message passing algorithms. This behavior suggests that most of the
solutions concentrate around the dominant point (the maximum in the
distance landscape) in the $N$-dimensional weight space. It is clear
that as the constraint density increases, the distance landscape
shows larger and larger deviation from the analytic annealed
approximation. We also calculate the second distance landscape
characterizing the number of solution-pairs separated by a given
distance. In this case, the replica symmetric result is in good
agreement with the annealed computation at low $\alpha$, while the
large deviation is observed between the replica symmetric
approximation and annealed computation for high $\alpha$. Both
landscapes are evaluated in the thermodynamic limit and confirmed by
message passing simulations on large-size single instances.

In this paper, we calculate the whole picture of the distance
(entropy) landscape and show that the entropy value rises to a
maximum before declining at higher values of distance at certain
range of distances. From the first landscape (a random configuration
as a reference), we see clearly how the solution space shrinks as
more constraints are added. From the second landscape of
solution-pairs, we deduce a picture in which each global minimum
(referred to as a canyon) is occupied by a single solution with zero
ground state energy, and is surrounded by local minima (referred to
as valleys) with positive energy~\cite{Lenka-2010prb}. This is also
known as the valleys-dominated energy
landscape~\cite{Lenka-2008Jphys}. The isolation of solutions implies
that one cannot expect to satisfy all constraints by flipping a few
synapses. The necessary number of synapses to be flipped should be
proportional to $N$. The distance between the isolated solutions
increases as the constraint density grows. This is the very reason
why some simple local search heuristics cannot find a solution at
high $\alpha$ or large $N$~\cite{Huang-2010jstat,Huang-2011epl} and
the critical $\alpha$ for the local search algorithm decreases when
the number of synapses increases. Simulated annealing process used
in Refs.~\cite{Patel-1993,Horner-1992a} suffers from a critical
slowing down when approaching a certain temperature, therefore, it
would be interesting to study this picture within a finite
temperature framework by focusing on the metastable states around
the isolated solutions. The structure of these states should also be
responsible for the algorithmic hardness. This issue will be
addressed in our future work.

The distance landscape evaluated here is very similar to the weight
enumerator in coding theory~\cite{Montanari-2004} and the method can
be extended to consider the landscape analysis for low-density
parity-check codes or code-division multiple access multiuser
detection problems~\cite{Kaba-2003a,Kaba-2004}, which will help to
clarify what role the distance landscape plays with respect to the
decoding performance.


\section*{Acknowledgments}

We are grateful to Haijun Zhou for helpful comments on earlier versions of this paper. This work was
partially supported by the Research Council of Hong Kong (Grant Nos.
HKUST $605010$, $604512$)(MW, HH) and the JSPS Fellowship for Foreign Researchers
Grant No. $24\cdot02049$ (HH) and JSPS/MEXT KAKENHI Nos. $22300003$, $22300098$, and $25120013$ (YK).
\appendix

\section{Derivation of $\frac{{\rm d}S(R)}{{\rm
d}R}$ in the limit of $R\rightarrow1$} \label{sec:appendix-a} In
this section, we give a derivation of $\frac{{\rm d}S(R)}{{\rm d}R}$
in the limit of $R\rightarrow1$. By noting that $s(R)=f(x)-xR$, one
can write the derivative as
\begin{equation}\label{derR}
    \frac{{\rm d}S(R)}{{\rm d}R}=\frac{\alpha}{1-R}\int Dt\frac{\int Dy G^{2}(\tilde{y})}{\int Dy
    H^{2}(\tilde{y})}-\hat{R}
\end{equation}
where $\tilde{y}=-\frac{\sqrt{R-q}y+\sqrt{q}t}{\sqrt{1-R}}$ and
$\hat{R}$ is determined by Eq.~(\ref{pairR}). We know that
$R\rightarrow1$ implies that $\hat{R}\rightarrow\infty$, and let $\epsilon=1-R$ in Eq.~(\ref{pairR}), we can then obtain
\begin{equation}\label{derR01}
   \hat{R}=\hat{q}+\frac{1}{2}\ln\left(2\int
   Dz\frac{1-(\tanh\sqrt{\hat{q}}z)^{2}}{1+(\tanh\sqrt{\hat{q}}z)^{2}}\right)-\frac{1}{2}\ln\epsilon.
\end{equation}
To derive Eq.~(\ref{derR01}), we have used $\tanh(t)\simeq1-2e^{-2t}$ $(t\rightarrow\infty)$.
When $R\rightarrow1$, $\int Dy
G^{2}(\tilde{y})=\sqrt{\epsilon}\frac{G(-\frac{\sqrt{q}}{\sqrt{1-q}}t)}{2\sqrt{\pi(1-q)}}$
and $\int Dy H^{2}(\tilde{y})=H(-\frac{\sqrt{q}}{\sqrt{1-q}}t)$,
thus the first term in Eq.~(\ref{derR}) can be reexpressed in the
limit $R\rightarrow1$ as
\begin{equation}\label{derR02}
    \frac{\alpha}{1-R}\int Dt\frac{\int Dy G^{2}(\tilde{y})}{\int Dy
    H^{2}(\tilde{y})}=\frac{\alpha}{2\sqrt{\pi}}\int
    Dt\left[\frac{G(-\frac{\sqrt{q}}{\sqrt{1-q}}t)}{\sqrt{1-q}H(-\frac{\sqrt{q}}{\sqrt{1-q}}t)}\right]\times
    \epsilon^{-1/2}.
\end{equation}
Therefore, $\frac{{\rm d}S(R)}{{\rm d}R}=\alpha
C_{p}\epsilon^{-1/2}+C+\frac{1}{2}\ln\epsilon$ where the constants
can be obtained from the above equations. Note that $C_{p}$ is
positive, and the first term in the derivative dominates the
divergent behavior when $\epsilon\rightarrow0$. Following the same
line, one can also prove that $s(R)$ reduces to $s_{{\rm rs}}$ in
the limit of $R\rightarrow 1$.

\section*{References}


\end{document}